\documentclass[prb,preprint, superscriptaddress, amsmath, amssymb]{revtex4-1}
\usepackage{float} 
\usepackage{graphicx}
\usepackage{dcolumn}
\usepackage{bm}
\usepackage{upgreek}
\usepackage{times}
\usepackage{mathrsfs}
\usepackage{multirow}
\usepackage{hyperref} 
\usepackage{fancyheadings}
\usepackage[version=3]{mhchem}
\usepackage{braket}

\RequirePackage[normalem]{ulem}
\RequirePackage{color}\definecolor{RED}{rgb}{1,0,0}\definecolor{BLUE}{rgb}{0,0,1}

\begin{document}

\title{Exploration of the bright and dark exciton landscape and fine structure of MoS$_{2}$ (using G$_0$W$_0$-BSE)}
\author{Hongyu Yu}
\affiliation{Department of Materials Science and Engineering, McMaster University, 1280 Main Street West,
Hamilton, Ontario L8S 4L7, Canada}
\affiliation{School of Physics, Nankai University, Tianjin 300071, China}
\author{Magdalena Laurien}
\email{laurienm@mcmaster.ca}
\affiliation{Department of Materials Science and Engineering, McMaster University, 1280 Main Street West,
Hamilton, Ontario L8S 4L7, Canada}
\author{Zhenpeng Hu}
\affiliation{School of Physics, Nankai University, Tianjin 300071, China}
\author{Oleg Rubel}
\email{rubelo@mcmaster.ca}
\affiliation{Department of Materials Science and Engineering, McMaster University, 1280 Main Street West,
Hamilton, Ontario L8S 4L7, Canada}

\date{\today}
\begin{abstract}
Spectral ordering between dark and bright excitons in transition metal dichalcogenides is of increasing interest for optoelectronic applications. However, little is known about dark exciton energies and their binding energies. We report the exciton landscape including momentum-forbidden dark excitons of MoS$_{2}$ monolayer using single shot GW-Bethe Salpeter equation (G$_{0}$W$_{0}$-BSE) calculations. We find the lowest-energy exciton to be indirect at ($\textrm K'_{v} \rightarrow \textrm K_{c}$) in agreement with recent GdW-BSE calculations [2D Mater. 6, 035003 (2019)]. We also find that by large, the lowest-energy dark exciton binding energies ($E_b$) scale with the quasiparticle energies ($E_g$) according to the empirical $E_b/E_g=0.25$ rule. Differences in exciton binding energies are explained using an orbital theory. 
\end{abstract}


\maketitle
%
%
\section{Introduction}
Two-dimensional transition metal dichalcogenides (TMDCs) like monolayer (ML) MoS$_{2}$ exhibit an intricate electronic fine structure that offers an abundance of possibilities to manipulate their optical and electrical properties and exploit them for novel devices. A fascinating aspect of ML materials that sets them apart from their bulk equivalents is the behavior of excitations: Quasi-particles formed by an excited electron and a hole (excitons) experience a greater Coulomb attraction in a monolayer material because of the lack of screening in the third dimension. These excitonic effects dominate the optical response of ML TMDCs.

Excitons can be either bright (optically accessible) or dark (optically inaccessible). Dark excitons can be classified according to two main characteristics: spin and location in momentum space of the electron and hole. Spin-forbidden dark excitons are quasiparticles where the electron and the hole occupy the same position in momentum space, however, their spin is opposite and thus radiative recombination is not possible. Momentum-forbidden dark excitons consist of an electron and a hole located at different points in momentum space. Unassisted recombination is not possible for these indirect excitons either, thus they are dark.

Besides the bright states, dark excitons have a considerable influence on the optical response of TMDC MLs.\cite{mueller2018exciton}  For example, spectral closeness of dark excitons to bright excitons can cause a significant drop of the photoluminescent yield in ML MoS$_{2}$ \citep{wu2015exciton}. Similarly, higher-energy momentum-forbidden dark excitons can serve as a reservoir of charge carriers for bright transitions that are lower in energy and thus enhance the response for TMDC MLs \citep{steinhoff2017exciton}. Indirect excitons have also been related to the achievable degree of circular polarization in TMDC MLs \citep{baranowski2017dark} and the formation of quantum dots in bilayer WSe$_{2}$ \citep{lindlau2018role}. In addition, dark excitons in WSe$_2$ can be activated or brightened, i.e. the photoluminescence intensity increases, in the presence of a magnetic field which leads to the creation of bright excitons with long and tunable life times.\citep{zhang2017magnetic, vasconcelos2018dark} Brightening can also be achieved by strain \citep{feierabend2018dark} or the adsorption of high-dipole molecules \citep{feierabend2017proposal}, allowing for completely new device concepts in the design of high-sensitivity sensors.

Knowledge of the spectral relation of dark and bright excitons is important to fully understand the optical response of monolayer TMDCs.\citep{mueller2018exciton, selig2018dark} This is especially crucial for ML MoS$_{2}$ for which the ordering of the lowest-energy bright and dark excitons is still being discussed \citep{qiu2013optical, echeverry2016splitting, molas2017brightening}. The spectral ordering of bright and dark excitons depends mainly on the amount of band splitting caused by spin-orbit coupling as well as difference of the exciton binding energies. Initially, the emphasis was placed on studying direct excitons.\citep{wang2018colloquium} However, comprehensive quantitative studies of the excitonic landscape including indirect, finite-momentum excitons are scarce. Important contributions were made first by \citet{malic2018dark} who calculated the optical response of group-VI TMDCs. They emphasized the importance of excitonic corrections to the band structure that can lead to a change of the band character from direct to indirect or affect the ordering of bright and dark states. However, their results showed only qualitative trends. \citet{berghauser2018mapping} obtained the exciton landscape of monolayer MoS$_{2}$ and other group-VI TMDCs using pump-probe experiments and an empirically parameterized quantum model. According to their study, the lowest-energy state for ML MoS$_{2}$ is a dark (indirect) exciton with its hole located at $\Gamma_{v}$ and the electron located at $\textrm K'_{c}$ ($\Gamma_{v} \rightarrow \textrm K'_{c}$). Very recently, \citet{deilmann2019finite} reported calculations of the exciton landscape including indirect excitons in the GdW+BSE scheme, where the approximation $dW=W - W_\text{metal}$ enables a higher computational efficiency\citep{druppel2018electronic}. They found the excitonic state of monolayer Mo$X_{2}$ to be dark (indirect) and located at $\textrm K_{v} \rightarrow \textrm K'_{c}$.

In this report, we use \textit{ab initio} calculations [single-shot GW (G$_0$W$_0$) + BSE beyond the Tamm-Dancoff approximation (TDA)] to explore the whole bright and dark excitonic landscape of ML MoS$_{2}$ to contribute to the ongoing discussion. According to our results, the exciton ground state is a dark indirect exciton at $\textrm K_{v} \rightarrow \textrm K'_{c}$. We show that lowest-energy spin-forbidden and indirect excitons obey the universal relationship between exciton energy and exciton binding energy proposed for bright excitons in ML 2D materials\citep{jiang2017scaling}. We discuss the variations in the binding energies in the light of orbital theory. We also show that the relationship breaks down for higher-energy excitons.

%
%
\section{Methods}
\subsection{G$_0$W$_0$+BSE calculations}
We performed G$_0$W$_0$-BSE \textit{ab initio} calculations. The procedure for GW-BSE calculations is as follows: In the GW step the electronic ground state previously obtained using density functional theory \cite{Kohn_PR_140_1965} is corrected for quasiparticle effects. This correction is obtained by solving for the self-energy which includes the many-body exchange-correlation interactions in a single shot. In Hedin's method \citep{hedin1965new}, the self-energy is approximated by the product of the one-particle Green's function $G$ and the screened Coulomb potential $W$. The quasiparticle corrected energies and wave functions are used as input for the BSE which describes interactions of electron-hole pairs and directly yields the optical excitation energies. The exciton wave function is constructed as an expansion in terms of quasiparticle wave functions, and then the BSE can be solved self-consistently as an eigenvalue problem. In most cases, it is sufficient to solve the BSE in the TDA \citep{Tamm1991, Dancoff1950, leng2016gw}:
\begin{equation}\label{Equ:bse_tda}
\left(E_{c \mathbf{k}+\mathbf{Q}}-E_{v \mathbf{k}}\right) A_{v c \mathbf{k}}^{(S, \mathbf{Q})}+\sum_{v^{\prime} c^{\prime} k^{\prime}} K_{v c \mathbf{k}, v^{\prime} c^{\prime} \mathbf{k}^{\prime}}^{A A}(\mathbf{Q}) A_{v^{\prime} c^{\prime} \mathbf{k}^{\prime}}^{(S, \mathbf{Q})}{\quad=\Omega^{(S, \mathbf{Q})} A_{v c \mathbf{k}}^{(S, \mathbf{Q})}} .
\end{equation}
Here $\Omega^{(S, \mathbf{Q})}$ is the exciton energy (the eigenvalue), $E_{v \mathbf{k}}$ ($E_{c \mathbf{k}+\mathbf{Q}}$) are the energies of the valence band (conduction band) obtained in the GW step, $A_{v c \mathbf{k}}^{(S, \mathbf{Q})}$ are expansion coefficients for the exciton wave function, and $K$ is the interaction kernel which contains all the electron-hole interactions. Details concerning the mathematical form of $K$ can be found in \citet{leng2016gw}. The index $\mathbf{Q}$ denotes a momentum transfer by a certain $\mathbf{Q}$ vector. Here, we went beyond the TDA, including resonant-antiresonant coupling ($K^{AB}$, $K^{BA}$):\citep{leng2016gw} 
\begin{equation}\label{Equ:bse_notda_test}
\left\{\begin{aligned}
\left(E_{c \mathbf{k}+\mathbf{Q}}-E_{v \mathbf{k}}\right) A_{v c \mathbf{k}}^{(S, \mathbf{Q})}+\sum_{v^{\prime} c^{\prime} k^{\prime}} K_{v c \mathbf{k}, v^{\prime} c^{\prime} \mathbf{k}^{\prime}}^{A A}(\mathbf{Q}) A_{v^{\prime} c^{\prime} \mathbf{k}^{\prime}}^{(S, \mathbf{Q})} \\ +\sum_{v^{\prime} c^{\prime} k^{\prime}} K_{v c \mathbf{k}, v^{\prime} c^{\prime} \mathbf{k}^{\prime}}^{A B}(\mathbf{Q}) B_{v^{\prime} c^{\prime} \mathbf{k}^{\prime}}^{(S, \mathbf{Q})}{\quad=\Omega^{(S, \mathbf{Q})} A_{v c \mathbf{k}}^{(S, \mathbf{Q})}} \\
\left(E_{c \mathbf{k}+\mathbf{Q}}-E_{v \mathbf{k}}\right) B_{v c \mathbf{k}}^{(S, \mathbf{Q})}+\sum_{v^{\prime} c^{\prime} k^{\prime}} K_{v c \mathbf{k}, v^{\prime} c^{\prime} \mathbf{k}^{\prime}}^{B B}(\mathbf{Q}) B_{v^{\prime} c^{\prime} \mathbf{k}^{\prime}}^{(S, \mathbf{Q})} \\ +\sum_{v^{\prime} c^{\prime} k^{\prime}} K_{v c \mathbf{k}, v^{\prime} c^{\prime} \mathbf{k}^{\prime}}^{B A}(\mathbf{Q}) A_{v^{\prime} c^{\prime} \mathbf{k}^{\prime}}^{(S, \mathbf{Q})}{\quad=\Omega^{(S, \mathbf{Q})} B_{v c \mathbf{k}}^{(S, \mathbf{Q})}}
\end{aligned}\right.
\end{equation}
Where $B_{v c \mathbf{k}}^{(S, \mathbf{Q})}$ are expansion coefficients for the antiresonant part of the exciton wave function.

The main reason to conduct calculations beyond the TDA for our work was that the software used does not recommend the calculation of finite-momentum excitons within the TDA. The TDA has been shown to break down for nanoscale systems \citep{ma2009excited,puschnig2013excited,gruning2009exciton} and to deviate from experiment for finite-momentum excitons in silicon.\cite{sander2015beyond} However, we do not expect the resonant-antiresonant coupling to have a great effect on the optical properties of a ML TMDC.

To the best of our knowledge, solving the BSE beyond the TDA has not yet been reported for group-VI metal transition dichalcogenides. In the following we describe the details of our settings used to perform the G$_0$W$_0$-BSE beyond TDA calculations.

\subsection{Computational details}
The calculations were performed with the Vienna ab initio package (VASP) \cite{kresse1993ab, kresse1996efficiency}, version 5.4.4. The projector-augmented wave method \citep{blochl1994projector, kresse1999ultrasoft} was used to treat core and valence electrons with 14 electrons for Mo, and 6 electrons for S explicitly included in the valence states. The plane-wave energy cutoff was set to 400 eV. Recommended GW projector-augmented wave potentials supplied by VASP were employed for all atoms. The Perdew-Burke-Ernzerhof\citep{perdew1996generalized} exchange-correlation functional was used to obtain the electronic ground state with density functional theory\citep{hohenberg1964inhomogeneous, kohn1965self}. To ensure minimal interlayer coupling, monolayers were separated by 21.5~{\AA} of vacuum which is sufficient for the longitudinal component of the macroscopic static dielectric tensor to be close to unity. Atomic positions and lattice vectors were fully relaxed with a tolerance of 0.01~eV/{\AA}. Only the $c$ vector (out-of-plane vector) was fixed during the relaxation procedure. Electronic minimization was performed with a tolerance of 10$^{-7}$~eV and convergence accelerated with Gaussian smearing of the Fermi surface by 0.05~eV. The Brillouin zone was sampled with a $12 \times 12 \times 1$ $\Gamma$-centered $k$-point mesh in order to include high symmetry points in the $k$ mesh and ensure sufficient accuracy of the exciton binding energy that is highly dependent on the density of the $k$ mesh\citep{bokdam2016role}. After structure relaxation, we obtained a lattice constant of 3.185~{\AA}, a metal-chalcogen (M-X) bond length of 2.414~{\AA}, and a chalcogen-chalcogen X-X bond length of 3.12838~{\AA}. The obtained lattice constant is close to the experimental lattice constant of bulk MoS$_{2}$ ($a = 3.16$~{\AA}) \citep{bronsema1986structure, schonfeld1983anisotropic, wildervanck1964preparation} and in excellent agreement with other computational studies\citep{kang2013band, liu2013three, ramasubramaniam2012large}. The M-X bond length is in very good agreement with experimental data \citep{samadi2018group, dickinson1923crystal}.

For all calculations following the relaxation procedure, we considered spin-orbit coupling and included 640 bands (26 of them occupied) in order to have enough empty bands for the ensuing GW calculations. Further, the orbitals were enforced to have real values at the Gamma point and points at the edge of the Brillouin zone and as a consequence the symmetry was turned off. 

We calculated the quasiparticle band structure at the single-shot G$_0$W$_0$ level of theory. For the response function we set a cutoff of 250~eV; this parameter controls how many G-vectors are included in the GW-calculation. The number of frequency grid points was set to 96. For visualizing the quasiparticle band structure we applied Wannier interpolation using the WANNIER90 program\citep{mostofi2008wannier90}. 

The BSE calculations were carried out beyond the Tamm-Dancoff approximation using the full BSE Hamiltonian \citep{sander2015beyond}, which means that resonant-antiresonant coupling is included. For solving the BSE, we considered 6 occupied bands and 8 virtual (unoccupied) bands of the quasiparticle band structure as a basis for excitonic eigenstates. To obtain finite-momentum excitons, we iterated over all possible $\mathbf{Q}$ vectors that could be selected for the given $k$ mesh in the first Brillouin zone (in total 144) and additional  $\mathbf{Q}$ vectors outside the first Brillouin zone to include the  $\textrm K'_{v} \rightarrow \textrm K_{c}$,  $\textrm K_{v} \rightarrow \textrm K'_{c}$ and  $\textrm K'_{v} \rightarrow \Lambda_{c}$ transitions. For all $\mathbf{Q}$ vectors, we obtained the lowest 100 eigenstates as output. We chose the $k$ point with the biggest contribution to the exciton wave function (highest amplitude) for each eigenstate as the momentum vector of the hole of the exciton. Exciton binding energies were calculated by subtracting the BSE eigenvalues from the GW band gap matching the position of hole and electron of the exciton in momentum space. To distinguish between spin-parallel and spin-antiparallel states, the spinor up and down components ($\alpha$ and $\beta$) were determined from spin projections as described in Refs.~\citenum{giustino2014materials, zheng2018structural}. 


We would like to point out certain limits of our methods. The G$_0$W$_0$-BSE procedure as implemented in VASP and as used for this work does not provide the option to truncate the Coulomb interaction between periodic images. Carefully conducted studies\citep{huser2013dielectric, qiu2016screening} show that Coulomb truncation is essential for achieving convergence of the GW band energy corrections, as without the truncation the periodic images of the monolayer increase the dielectric function, especially in the low $\mathbf{Q}$ limit. Further, a very high $k$ mesh up to $ 300 \times 300 \times 1$ is required in order to converge the exciton binding energy to within 0.1~eV.\citep{qiu2016screening} As our G$_0$W$_0$ calculations and the BSE beyond TDA calculations were conducted without considering geometrical and time reversal symmetries, the computational cost precludes the use of fine $k$ meshes (due to excessive memory requirements). However, the errors of not truncating the Coulomb interaction and using a coarse $k$ mesh partly cancel out.\citep{huser2013dielectric} 

We conducted convergence tests that suggest that the total error of the quasiparticle band gap is below 0.1 eV and the variation of the spectral spacing with $k$ grid density is ca. one order of magnitude smaller than the actual energy spacing \cite{si_convergence}.
%
%
\section{Results and discussion}\label{Sec:Results and discussion}



We will first discuss the effect of including the resonant-antiresonant coupling (going beyond the TDA). The TDA affects only the BSE step of the calculations. We performed a comparative BSE calculation employing the TDA to investigate the effects of the absence of resonant-antiresonant coupling on the optical properties of ML MoS$_{2}$. The results are identical to the full BSE calculations, differences are negligible. This is true for the spectral spacing of the excitons \cite{si_TDA} as well as for the dielectric response (data not shown). 

\begin{figure}[h]
	\includegraphics{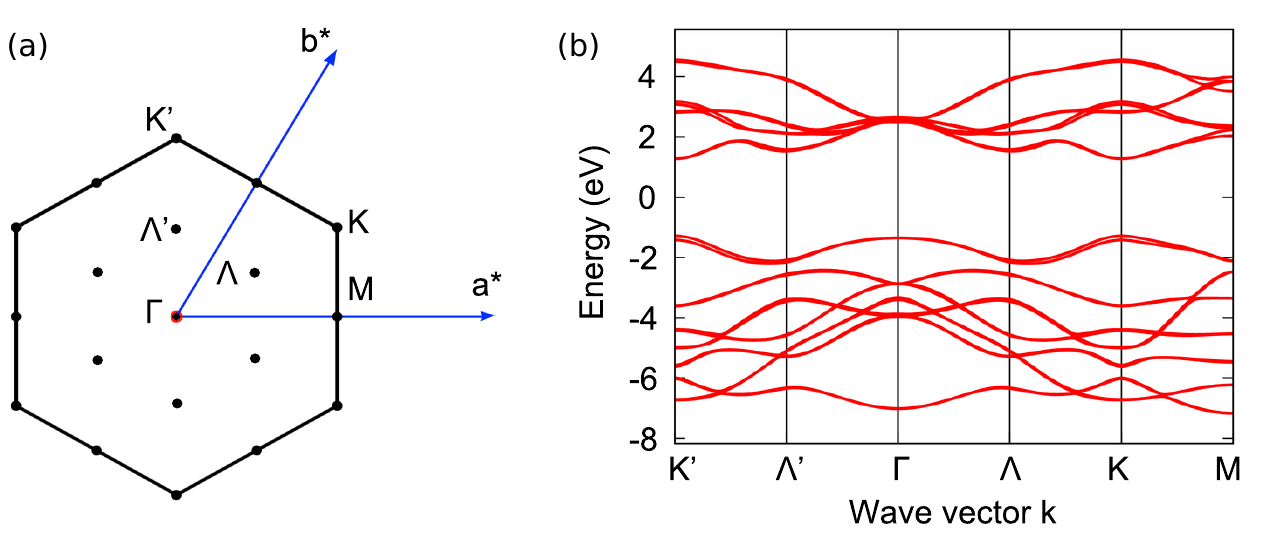}
	\caption{(a) First Brillouin zone of monolayer MoS$_{2}$ and (b) quasiparticle  band structure after Wannier-interpolation. The direct band gap is located at K ($\textrm K'$).}
	\label{Fig:BZ_Wannier}
\end{figure}

Now we turn to the results of the main calculations. The quasiparticle band structure of monolayer MoS$_{2}$ is shown in Fig.~\ref{Fig:BZ_Wannier}. The bands are obtained by Wannier interpolation of the GW eigenvalues. The band structure shows a direct band gap of ca. 2.43 eV at the K and $\textrm K'$ points. These points are equivalent (except for their spin) because of the time-reversal symmetry. Besides K and $\textrm K'$, $\Lambda$ and $\Lambda'$ are related via time-reversal symmetry. For future discussions we will only refer to one of the via time-reversal symmetry related transitions. 

\begin{figure}[h]
	\includegraphics{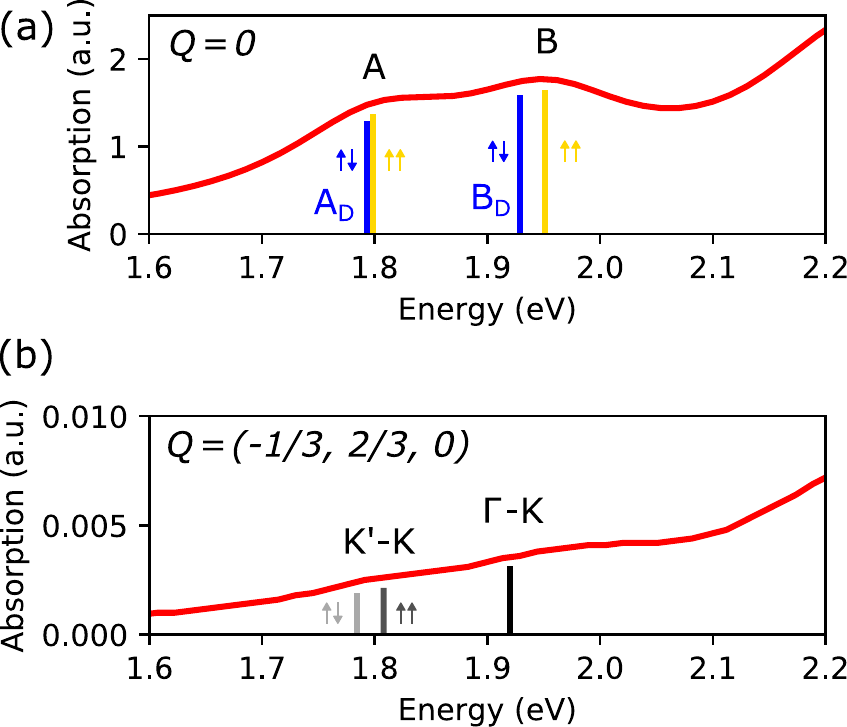}
	\caption{Absorption spectra of MoS$_{2}$. Shown is the imaginary part of the dielectric function obtained from the BSE calculation. (a) $\mathbf{Q}=0$ denotes the absorption spectrum of the direct transitions. The first two absorption peaks denote the A and B excitons located at the K point. The dark (spin-forbidden) exciton is lower in energy than the bright exciton for both A and B excitons. (b) $\mathbf{Q}=(-1/3, 2/3, 0)$ captures the the $\textrm K'_{v} \rightarrow \textrm K_{c}$ and $\Gamma_{v} \rightarrow \textrm K'_{c}$ transitions. The indirect exciton at $\textrm K'_{v} \rightarrow \textrm K_{c}$ is clearly lower in energy than the dark direct exciton at K.}
	\label{Fig:spectra}
\end{figure}

The optical response of TMDC monolayers is dominated by the presence of excitons and their binding energies. As a result, the optical energy gap is much smaller than the quasiparticle band gap. This can be seen when considering the absorption spectrum of ML MoS${_2}$ for the direct transitions ($\mathbf{Q}=0$) obtained from our BSE calculations (Figure ~\ref{Fig:spectra}a). The A and B excitons are located at ca. 1.8 eV and 1.95 eV, implying binding energies of about 0.62 and 0.48 eV, respectively. The spin-forbidden dark exciton is slightly lower in energy than the bright exciton. The absorption spectrum is in good qualitative agreement with experiments\citep{mak2010atomically, dhakal2014confocal} as well as other theoretical studies\citep{palummo2015exciton, bernardi2017optical}. The energies of the A and B peak are blue-shifted in comparison to experiment. This comes about for  two reasons; missing substrate effects\citep{mukherjee2015complex} as well as $k$ grid dependent binding energies. The denser the $k$ grid, the smaller (i.e., better converged) the binding energies become \cite{bokdam2016role,si_kgrid}.

The values obtained from our calculations for the band gap and exciton binding energies are in good agreement with experiment. Using scanning tunneling spectroscopy and optical reflectance contrast measurements for MoS$_{2}$ on fused silica, \citet{rigosi2016electronic} obtained a binding energy of the bright excitons of $0.31 \pm 0.04$~eV and an electronic band gap of $2.17 \pm 0.1$~eV. The results of our calculation with $E_b = 0.624$~eV and $E_g = 2.42$~eV are slightly higher than the experimental values because the calculations are obtained for a free-standing monolayer and a relatively coarse $k$ mesh. Other computational studies using GW-BSE found results that are quite close to ours obtaining 2.42 eV for the band gap and 0.57 eV for bright exciton binding energy \citep{jiang2017scaling}. The difference in the Eb can be explained with the $k$ mesh density: \citet{jiang2017scaling} used a $k$ mesh of $16 \times 16 \times 1$ (our calculations: $12 \times 12 \times 1$) and the binding energy strongly depends on the $k$ mesh \citep{bokdam2016role}. For example, in our convergence calculations for the lowest-energy direct exciton at the K point, we found an exciton binding energy of 0.551 eV for a $k$ grid of $15 \times 15 \times 1$ and nearly a twice as large binding energy of 1.061 eV for a $k$ grid of $6 \times 6 \times 1$ \cite{si_kgrid}.

Figure ~\ref{Fig:spectra}b  shows the spectra for indirect excitons with a $\mathbf{Q}$ vector of (-1/3, 2/3, 0). This $\mathbf{Q}$ vector captures the the $\textrm K'_{v} \rightarrow \textrm K_{c}$ and $\Gamma_{v} \rightarrow \textrm K'_{c}$ transitions. It becomes clear that there exists a smaller-energy exciton that is indirect at $\textrm K'_{v} \rightarrow \textrm K_{c}$ with an exciton energy of less than 1.8 eV.

\begin{figure}[h]
	\includegraphics{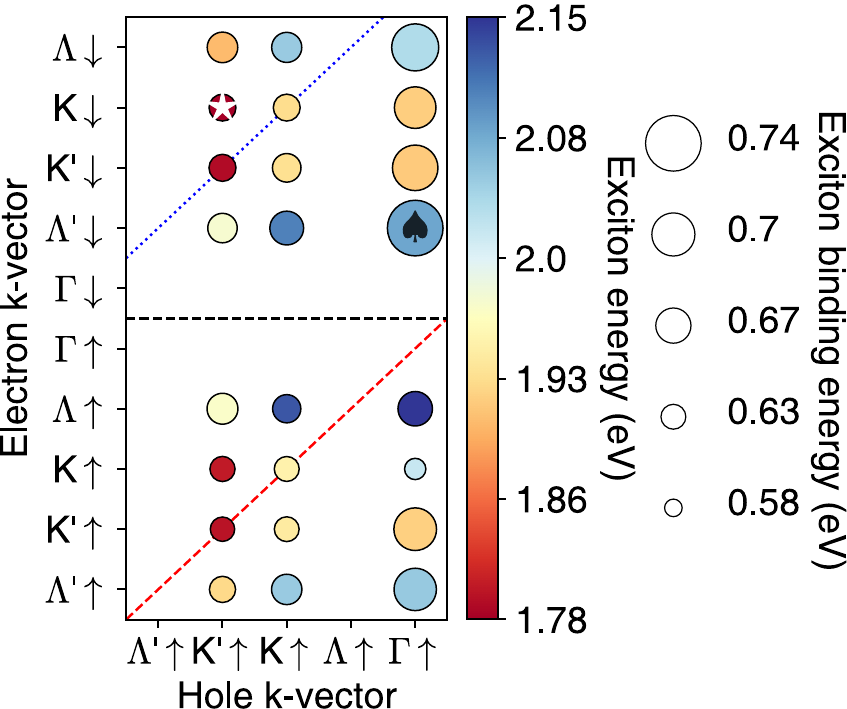}
	\caption{MoS$_{2}$ exciton landscape including dark and bright excitons. The color scale reflects the exciton energy and the circle size represents the exciton binding energy. In order to display finite-momentum excitons, hole and electron $k$ vectors are displayed separately on the horizontal and vertical axis, respectively. The lowest-energy exciton (star) is located at $\textrm K'_{v}\uparrow \rightarrow \textrm K_{c}\downarrow$ and is therefore momentum-forbidden. The highest binding energy (spades) occurs for an indirect exciton at $\Gamma_{v}\uparrow \rightarrow \Lambda'_{c}\downarrow$.}
	\label{Fig:EBEG}
\end{figure}

To capture the effect of all important indirect excitons on the quasiparticle band structure, we plot the exciton band structure in a two-dimensional fashion. This allows us to show the renormalization of the eigenvalues caused by direct, indirect, and dark excitons at the same time. In Fig.~\ref{Fig:EBEG} the landscape of bright and dark excitons in ML MoS$_{2}$ is shown for the most important points in momentum space. To accommodate momentum forbidden dark excitons, the $k$ vectors of the electron and hole of an exciton are displayed separately on the two axes of the graph. Further, we distinguish between spin-up and spin-down states to allow for the visualization of spin-forbidden excitons. As a result, bright excitons are seen on the dashed red diagonal line, spin-forbidden direct excitons are on the dotted blue line, and momentum-forbidden excitons are located to the sides. Also, spin-allowed excitons are distributed in the lower half of the plot while spin-forbidden excitons being placed in the upper half. Each bubble represents an exciton; the colour displays the exciton energy and the radius of the bubble corresponds to the exciton binding energy. The symmetry of the wave function $\Phi$ of the exciton can be expressed as:
\begin{equation}
\Phi\left(\mathbf{k}_{h},\mathbf{k}_{e},\mathbf{s}_{h},\mathbf{s}_{e}\right)=\Phi^*\left(\mathbf{-k}_{h},\mathbf{-k}_{e},-\mathbf{s}_{h},-\mathbf{s}_{e}\right).
\end{equation}
As a result, excitons $\ket {\mathbf{k}_{h},\mathbf{k}_{e},\mathbf{s}_{h},\mathbf{s}_{e}}$ and $\ket {-\mathbf{k}_{h},-\mathbf{k}_{e},-\mathbf{s}_{h},-\mathbf{s}_{e}}$ should have the same properties. As necessitated by our procedure, we calculated the whole Brillouin zone irrespective of time reversal symmetry. Because of this, our results showed computational inaccuracies  in the single-digit meV range for the band energies (and  the K point exciton energies) between per definition of time reversal symmetry identical states. The energy values presented here are always chosen from the exciton with the lower energy of the two (by time reversal symmetry) identical states. Due to time reversal symmetry we show only one half of the hole states in Fig ~\ref{Fig:EBEG}. The other half would be equal to the first by center symmetry. Although unoccupied, we chose to include the $\Lambda_v$ and the $\Gamma_c$ states to preserve the center symmetry.
 
\begin{figure}[h]
	\includegraphics{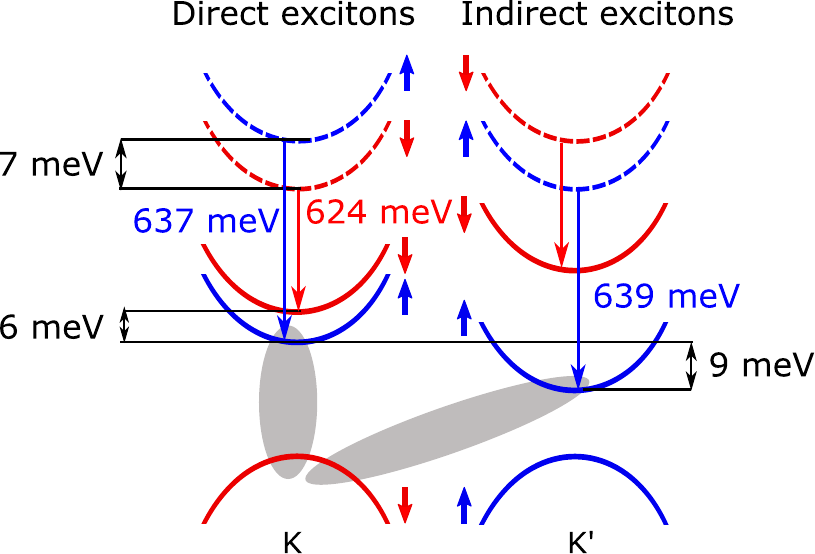}
	\caption{Schematic diagram of the lowest-energy excitons at the K-$\textrm K'$ valley of ML MoS$_2$. The ordering of solid energy bands corresponds to the exciton-corrected energies. The dashed conduction bands denote the quasiparticle band ordering. The lowest-energy exciton is indirect and located at K-K'. The lowest direct exciton is located at K and spin-forbidden (dark). This result is linked to spin-orbit coupling and the differences in the exciton binding energies which are higher for the spin- and momentum-forbidden excitons than for the bright exciton (637, 639, and 624 meV, respectively).}
	\label{Fig:exciton_valley}
\end{figure}

The exciton with the largest binding energy of 0.712 eV (marked with a spade) is located at $\Gamma_{v}\uparrow$ (hole) and $\Lambda'_{c}\downarrow$ (electron) ($\Gamma_{v}\uparrow \rightarrow \Lambda'\downarrow$). The lowest-energy exciton (marked with a star) has an energy of 1.784 eV and is located at $\textrm K'_{v}\uparrow \rightarrow \textrm K_{c}\downarrow$. This implies that after considering excitonic effects, we find a change of the optical band gap location of MoS$_{2}$  with regards to the transport band gap: the optical band gap is now indirect. The exciton at  $\textrm K'_{v}\uparrow \rightarrow \textrm K_{c}\downarrow$ is 15 meV lower in energy than the bright exciton at K and 9 meV lower in energy than the spin-forbidden direct exciton at K. The band ordering is illustrated in Fig.~\ref{Fig:exciton_valley}.

\begin{figure}[h]
	\includegraphics{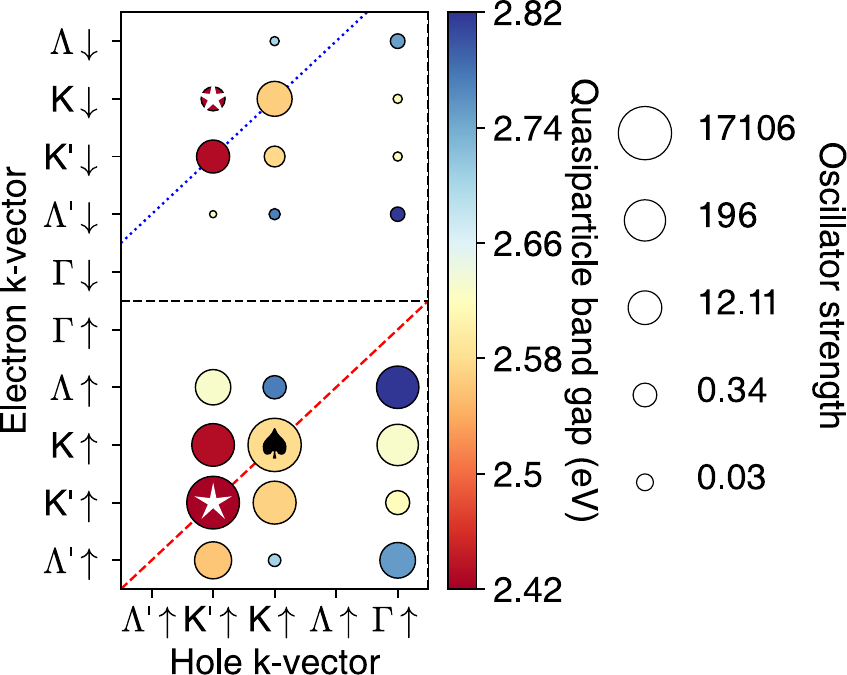}
	\caption{MoS$_{2}$ quasiparticle band gaps (color scale) and oscillator strengths (bubbles in a log scale) for bright and dark excitons. The quasiparticle band gap (star) is located at  $\textrm K'\uparrow$ ($\textrm K\downarrow$). An indirect band gap of equal energy is located at $\textrm K'_{v}\uparrow$ - $\textrm K_{c}\downarrow$. The oscillator strength of the bright transitions is several magnitudes larger than that of the dark transitions; the highest oscillator strength corresponds to the $\textrm K'_{v}\uparrow \rightarrow \textrm K'_{c}\uparrow$ transition (spades).}
	\label{Fig:OSGW}
\end{figure}

Besides the evaluation of the spin-states, dark and bright excitons can be distinguished by their oscillator strength. The oscillator strength of bright excitons is several magnitudes higher than of dark excitons \cite{echeverry2016splitting}. In Fig.~\ref{Fig:OSGW} we show the oscillator strength (bubble size) obtained from solving the BSE paired with the quasiparticle band gap (colour map) in a similar fashion to the exciton landscape. We find that the direct, spin-allowed excitons at K and $\textrm K'$ are about 1400 times higher in oscillator strength than their spin-forbidden equivalents.

Figure~\ref{Fig:OSGW} also shows that our GW calculations predict a direct band gap at K with the valence and conduction band having the same spin. The lower-energy spin-forbidden excited state after considering excitonic effects arises due to different exciton binding energies ($E_b$) of the dark and the bright exciton: The $E_b$ of the indirect dark exciton ($E_b = 0.637$~eV) is about 13 meV higher than the binding energy of the bright exciton ($E_b = 0.624$~eV) while the spin splitting of the conduction band is only about 7~meV (see Fig.~\ref{Fig:exciton_valley}). Thus, after considering excitonic effects, the spin- transition at K is lower in energy than the spin-allowed transition. These results are in agreement with \citet{qiu2013optical} and \citet{deilmann2019finite} who also found that the dark exciton at K is lower in energy than the bright exciton. \citet{echeverry2016splitting}, using the GW-BSE method came to the opposite conclusion. \citet{qiu2016screening} attribute the differing results in the literature to different settings of the density functional theory, GW and BSE parameters.

Now we will discuss the binding energies of the whole exciton landscape in more detail. Most of the holes of the excitons locate at $\Gamma$ or K and the electrons of the excitons locate at the $\Lambda$ or K (see Fig.~\ref{Fig:EBEG}). Interestingly, the exciton binding energy of the excitons whose hole is at $\Gamma$ is almost always higher than of excitons whose hole is located at K. For bulk semiconductors this effect could be explained with the effective mass differences as holes at the $\Gamma$ point are heavier than holes at the K point \citep{olsen2016simple}. However, for the binding energies of 2D materials the effective mass does not play a significant role, provided the polarizability is large (which is the case for MoS$_{2}$).\citep{olsen2016simple, jiang2017scaling}

Further, it is well known that the high binding energy of 2D materials originates from the lack of screening in the third dimension. Hence, we expect one factor for the different binding energies to be differences in screening depending on the position of the electron in real space. To qualitatively compare the amount of screening experienced by different excitons, we performed an orbital analysis for the valence and conduction band states of each exciton. The basic idea is that electrons occupying orbitals pointing perpendicular to the layer experience less screening than electrons of orbitals confined within the plane of the monolayer. By convention, $z$ is taken as the out-of-plane axis. It is well known, that for monolayer MoS$_{2}$ $\Gamma_{v}$ exhibits high contributions of the Mo-$d_{z^{2}}$ and S-$p_{z}$ orbitals while the $\textrm K_{v}$ state is mainly composed of $d_{xy}$ orbitals.\citep{samadi2018group} For our calculations we find the $\Gamma_{v}$  state to consist of ca. 77 \% $d_{z^{2}}$ + 22 \% $p_{z}$  and the $\textrm K_{v}$ state of  41 \% $d_{xy}$ + 41 \% $d_{x^{2}-y^{2}}$. Thus we can expect excitons at $\Gamma_{v}$ to experience less screening and consequently have higher binding energies than excitons at $\textrm K_{v}$.

\begin{figure}[h]
	\includegraphics{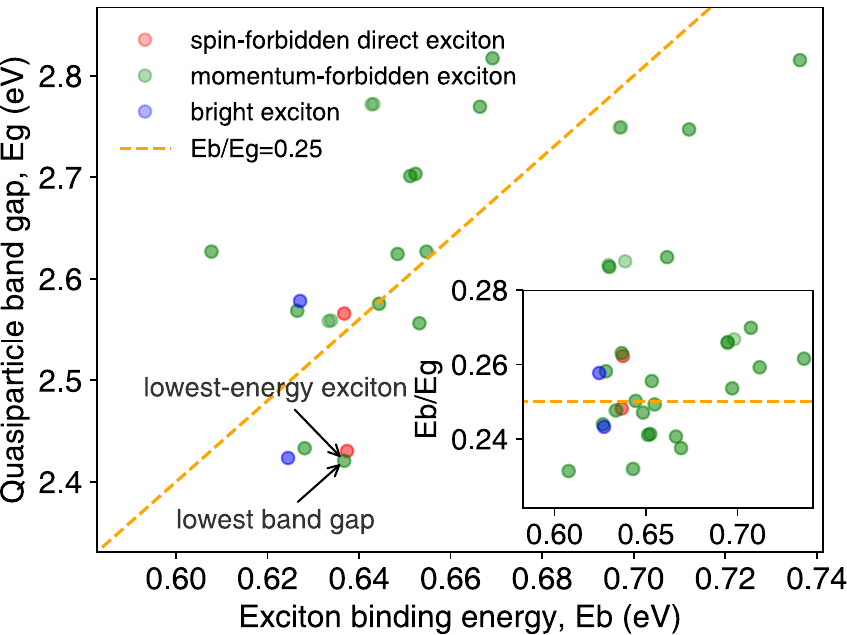}
	\caption{Ratio of the exciton binding energy ($E_b$) and quasiparticle band gap ($E_g$) for MoS$_{2}$. Bright excitons (blue), spin-forbidden excitons (red) and momentum-forbidden excitons (green) are shown. The  relationship of $E_b/E_b=0.25$ according to \citet{jiang2017scaling} is shown by a dashed line. The insert shows the ratio $E_b/E_g$ over $E_b$. All excitons are contained in the $E_b/E_g$ range of 0.23 - 0.28 and thus not to far from the 0.25 rule.}
	\label{Fig:relation}
\end{figure}

Figure~\ref{Fig:relation} shows the relationship between the quasiparticle band gap and exciton binding energy, including the lowest-energy bright, spin-forbidden and momentum-forbidden excitons. Generally, excitons at large quasiparticle band gaps have larger exciton binding energies. We included a dashed line in the figure that represents the $E_b/E_g = 0.25$ rule for excitons of 2D materials proposed by  \citet{jiang2017scaling}. Momentum-allowed excitons follow the rule that the exciton binding energy is about 0.25 of quasiparticle band gap \citep{jiang2017scaling} irrespective of their spin. Momentum-forbidden excitons also generally follow the trend of the exciton binding energy being about 0.25 of the band gap but with more scattering (ratios from 0.23 to 0.28). The insert of Figure~\ref{Fig:relation} shows a trend for indirect excitons: the higher the binding energy $E_b$, the higher is the ratio $E_b/E_g$. What causes this relationship?

\begin{figure}[h]
	\includegraphics{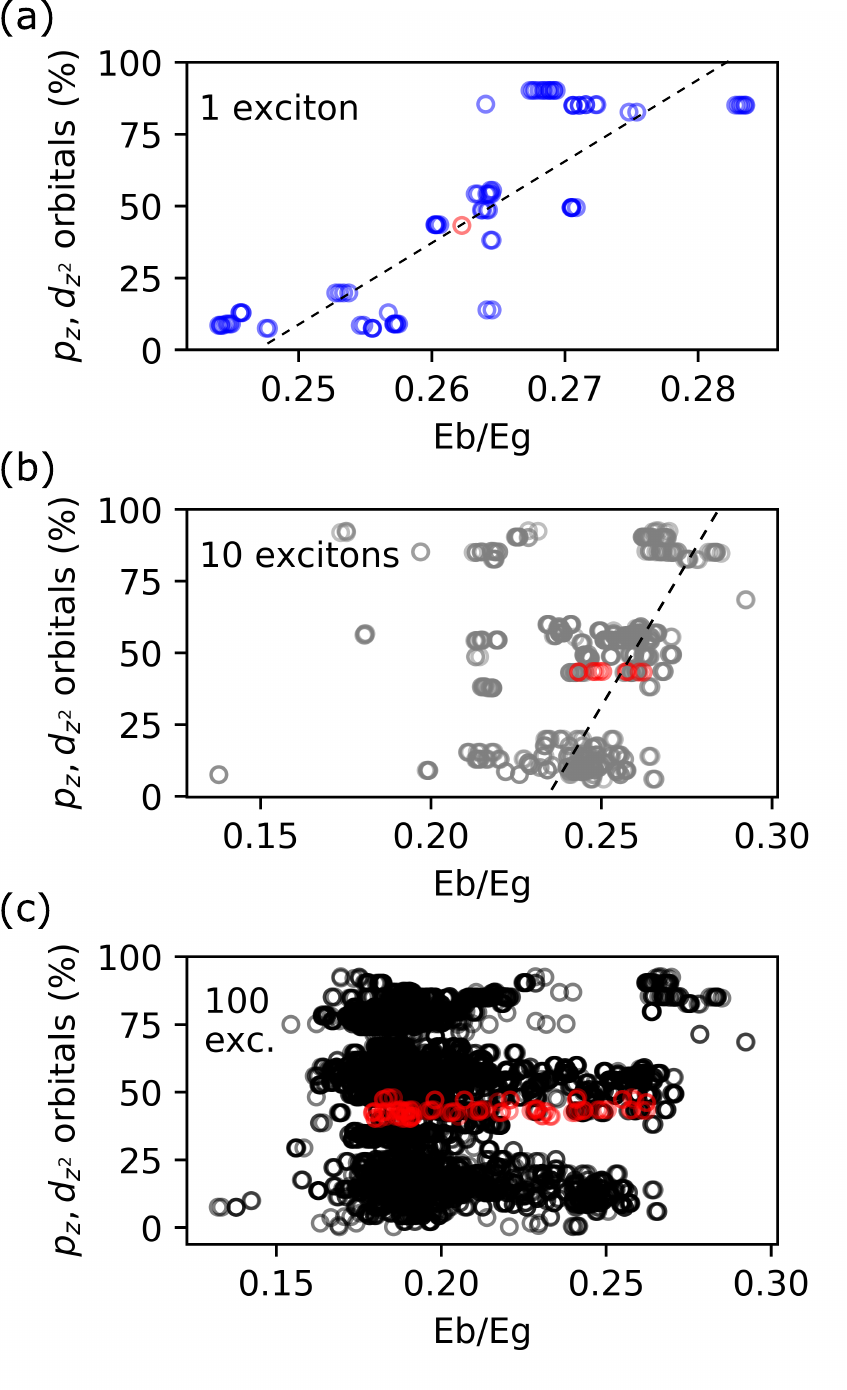}
	\caption{Orbital contributions to valence and conduction band states of indirect excitons in MoS$_2$ over the ratio of the exciton binding energy ($E_b$) and quasiparticle band gap ($E_g$) for MoS$_{2}$. Direct excitons are also shown (red). (a) For the lowest-energy excitons  (1 exciton per $\mathbf{Q}$ vector) there is a clear dependence of the $E_b/E_g$ ratio on the contributions of the p$_z$ and d$_{z^2}$ orbitals (a line is included as guide to the eye). (b), (c) The more higher-energy excitons are included for each $\mathbf{Q}$ vector, the less obvious this dependence becomes and finally vanishes (10 and 100 lowest-energy excitons). At higher energies, exciton charge carriers are screened and hole and electron progressively behave as free charge carriers causing the binding energy to decrease.}
	\label{Fig:relation_all}
\end{figure}

In order to explore this in more detail we will next consider the exciton landscape including higher-energy excitons, up to 100 per  $\mathbf{Q}$ vector, and their orbital compositions. We determined the $d_{z^{2}}$ and $p_{z}$ orbital contributions of the hole and electron states of each exciton and plotted the sum of them against the $E_b/E_g$ ratio (Figure~\ref{Fig:relation_all}). For the lowest-energy excitons, excitons with higher $E_b/E_g$ ratios show a higher percentage of  $d_{z^{2}}$ and $p_{z}$ orbitals. In other words, there exists a correlation between the $E_b/E_g$ ratio on the orbital contributions. This result explains the range in the $E_b/E_g$ ratio observed in Figure~\ref{Fig:relation}. It can also be seen that the direct excitons (red circles in Figure~\ref{Fig:relation_all}) are confined to a narrow region of $d_{z^{2}}$ and $p_{z}$ percentage, just below 50 \%. As a result, the $E_b/E_g$ ratio does not scatter as much for the direct lowest-energy excitons as for the indirect excitons. However, upon including higher-energy excitons, the $E_b/E_g = 0.25$ relationship completely breaks down; the $E_b/E_g$ ratio also becomes largely independent of the orbital composition. The relationship of binding energy and orbital contributions becomes less clear and has vanished when the 100 lowest-energy excitons for each  $\mathbf{Q}$ vector are included. We attribute this to the weaker electron-hole interactions for excitons with higher energies. At these energies, excitons are closely spaced and decrease rapidly in binding energy, behaving as uncorrelated electron-hole pairs.\cite{bernardi2017optical} The decrease of the $E_b/E_g$ ratio with quasiparticle energy is shown in the Supplemental Material \cite{si_ratio}.




%
%
\section{Conclusion}
In conclusion, we performed calculations of finite-momentum excitons in MoS$_{2}$ monolayer within and beyond the first Brillouin zone. It is found that the holes of the lowest-energy excitons are located at the $\Gamma$ or K valleys, while the electrons reside in the K or $\Lambda$ valleys. Our calculations predict the lowest-energy exciton to be indirect at K-$\textrm K'$ which is in agreement with recent GdW-BSE calculations \citep{deilmann2019finite}. The energy difference between the indirect exciton at K-$\textrm K'$ and the spin-forbidden direct exciton at K-K is about 9~meV. The bright exciton is located at K and 15~meV higher in energy than the lowest-energy exciton at K-$\textrm K'$.
We also discussed the exciton binding energies. The ratio of $E_b/E_g =0.25$ found for bright excitons in monolayer 2D materials holds true approximately for dark and indirect excitons. Excitons contained in orbitals that point out of plane and thus experience less local screening show higher binding energies. The relation of exciton binding energies to orbital composition and the  $E_b/E_g =0.25$ relation both break down for higher-energy excitons.

\appendix*

%
%
\begin{acknowledgments}
H.Y. acknowledges support by the National Science Fund of China for Talent Training in the Basic Sciences (No. J1103208). M.L. and O.R. acknowledge funding provided by the Natural Sciences and Engineering Research Council of Canada under the Discovery Grant Programs RGPIN-2015-04518.
Calculations were performed using a Compute Canada infrastructure supported by the Canada Foundation for Innovation under John R. Evans Leaders Fund.
\end{acknowledgments}

\section*{Author contributions}
H.Y. and M.L. contributed equally to this work.

%
%
\bibliographystyle{apsrev4-1}
\bibliography{refs}

%
%
\newpage


%
%
\clearpage

\end{document}